# Topological field-effect transistor with quantized ON/OFF conductance of helical/chiral dislocation states

Xiaoyin Li and Feng Liu*

Department of Materials Science and Engineering, University of Utah, Salt Lake City, Utah 84112, USA

Abstract

Topology is a key ingredient driving the emergence of quantum devices. Topological field-effect transistor (TFET) has been proposed to outperform the conventional FET by replacing the ON state with topology-protected quantized conductance, while the OFF state remains the same normal insulating characteristics and hence bears similar drawbacks. Here, we demonstrate a proof-of-concept TFET having both ON and OFF quantized conductance, by switching between helical and chiral topological screw dislocation (SD) states in three-dimensional topological insulators. A pair of SDs are configured with one acting as channel and the other as gate controlled by local magnetic field. A reversible field-switching is achieved with the ON and OFF conductance of $2e^2/h$ and $e^2/h$, respectively, as shown by tight-binding quantum transport calculations. Furthermore, $BaBiO_3$ is shown as a candidate material having the desired topological SD states, based on first-principles calculations. Our findings open a new route to high-fidelity topological quantum devices.

*Corresponding author.
fliu@eng.utah.edu

Topology and quantum coherence are two key ingredients driving the emergence of quantum devices with superior functionalities over the conventional electronic devices. Topological materials featured with topology-protected conducting quantum states are perfect candidates to improve device performance through quantized coherent charge and spin transport with minimized heat dissipation and robustness against disorder scattering [1-3]. Since the first proposals of quantum spin Hall insulators (QSHIs) [4,5], extensive theoretical and experimental efforts have been made to uncover high-quality topological materials [6-17] and disclose conceptually new quantum devices to take full advantage of the merits of topological properties [18,19].

FET is a basic building block of modern electronic devices, which operates by switching the electric signals between high and low conductivities, known as the ON and OFF states. Robust and precise ON and OFF signals and a high ON/OFF ratio are essential for high-performance FETs. Conventional FETs based on semiconductors usually experience multiple scatterings in transport of charge carriers, causing dissipation of energy and generation of waste heat, limiting the device performance. To overcome such constraints, TFET utilizing topological surface or edge states as conducting channels has been proposed, which are immune to external perturbations and non-magnetic disorders, to improve over the conventional counterparts [1,2,20,21]. To date, however, the reported TFETs generally exploit a topological metallic and a trivial insulating state with the former functioning as ON and the latter as OFF state [22-27], respectively. Consequently, although the ON state of such TFETs has topologically protected quantized conductance, its OFF state represented by a normal insulator suffering similar drawbacks as the conventional FETs, such as the disorder induced current leakage [28] resulting in a nonzero OFF signal and a decreased ON/OFF ratio. In order to produce high-fidelity signals for both states, it is highly desirable that the central building block of TFET to function with both ON and OFF quantized conductance in the topological regime.

In this Letter, we exploit the potential of topological dislocation states in three-dimensional (3D) topological insulators (TIs), and propose a conceptually new high-fidelity TFET with both quantized ON and OFF states. Dislocations are commonly seen in crystalline materials as line defects [29-31], which are usually thought detrimental to material properties [32-35]. However, recent studies have demonstrated the prospect of reinventing dislocations as a valuable resource, such as for spintronics applications [36,37]. Interestingly, dislocations in 3D TIs have been demonstrated to induce topological gapless states when their Burgers vector **b** and the bulk topological invariants satisfy certain rules [38]. Taking advantage of these existing knowledge, here we further demonstrate that a pair of screw dislocations (SDs) in 3D TIs can serve as 1D conducting channels harvesting helical topological gapless states lying in the bulk band gap, contributing a quantized conductance of $2e^2/h$. Moreover, one of the two pairs of conducting channels can be selectively gapped out by applying a local Zeeman field, with the other pair left intact contributing a quantized conductance of $e^2/h$ associated with the 1D chiral topological line states. Accordingly, we devise a proof-of-concept TFET as illustrated in Fig. 1, where a pair of SDs are configured with one acting as channel contacted with two metal leads, and the other as gate controlled by a local magnetic field generated by a gate voltage. The conducting channels of topological dislocation states are reversibly switched between helical and chiral states by the local magnetic field, constituting the ON and OFF states with quantized conductance of $2e^2/h$ and $e^2/h$ respectively. As the ON and OFF states are both topologically protected, the device is robust against disorders in the whole operation regime, as demonstrated directly by tight-binding (TB) quantum transport calculations. Also, the ON and OFF states have distinct integer conductance, leading to effectively an infinite ON/OFF ratio. Furthermore, as a major step towards practical applications, we propose a strong 3D TI, $BaBiO_3$ [39], as a candidate material to realize the desired tunable topological dislocation states.

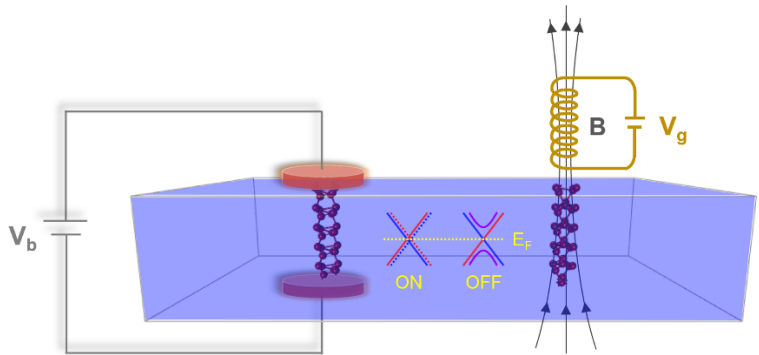

FIG. 1. Schematic illustration of the high-fidelity TFET enabled by topological dislocation states. The local leads, with a bias voltage $V_b$, cover one dislocation acting as quantized conducting channel. The magnetic field generated by the gate voltage $V_g$ is applied to the other dislocation, acting as a "gate" to switch between the quantized ON and OFF states. The inset displays the metallic helical/chiral state for the ON/OFF operation. $E_F$ is the Fermi level.

*Model.*–To illustrate the key principle of our proposal, the Fu-Kane-Mele model on a diamond lattice [40] hosting SDs is adopted for illustration. The TB Hamiltonian based on a single-orbital diamond lattice reads [40,41]:

$$H = t\sum_{\langle ij \rangle} c_i^\dagger c_j + i\lambda_{SO} \sum_{\langle\langle ij \rangle\rangle} c_i^\dagger \mathbf{s} \cdot \hat{\mathbf{e}}_{ij} c_j. \quad (1)$$

The first and second terms correspond to the nearest-neighbor (NN) hopping and second NN spin-orbit coupling (SOC), with $c^\dagger$ ($c$) being the electron creation (annihilation) operator. $\mathbf{s}$ denotes the vector of spin Pauli matrices and $\hat{\mathbf{e}}_{ij}$ is the unit vector defined by $\hat{\mathbf{e}}_{ij} = \left(\vec{d}_{ij}^{\,1} \times \vec{d}_{ij}^{\,2}\right)/\left|\vec{d}_{ij}^{\,1} \times \vec{d}_{ij}^{\,2}\right|$, where $\vec{d}_{ij}^{\,1}$ and $\vec{d}_{ij}^{\,2}$ are the two NN bond vectors connecting the second NN orbitals $j$ and $i$. Here we set $t = -1$, $\lambda_{SO} = -0.3$, and the NN hopping along the [111] direction is intentionally enhanced to ($t+\delta t$) with $\delta t = -1$, to drive the model into a strong TI phase having topological indices (1;111) [40]. All parameters are in the unit of eV. As shown previously [38], the dislocation will induce topological gapless states if its Burgers vector $\mathbf{b}$ and the time-reversal-invariant momentum (TRIM) $\mathbf{M}_\nu$ of the bulk material satisfy

$$\mathbf{b} \cdot \mathbf{M}_\nu = \pi \pmod{2\pi}. \quad (2)$$

The TRIM is defined by $\mathbf{M}_\nu = (1/2)(\nu_1 \mathbf{G}_1 + \nu_2 \mathbf{G}_2 + \nu_3 \mathbf{G}_3)$, where $(\nu_1, \nu_2, \nu_3)$ and $(\mathbf{G}_1, \mathbf{G}_2, \mathbf{G}_3)$ are the weak topological indices and the reciprocal lattice vectors respectively. Given a diamond lattice with lattice vectors $(\mathbf{a}_1, \mathbf{a}_2, \mathbf{a}_3)$, we introduce a pair of separated SDs with the opposite Burgers vectors $\mathbf{b} = \pm\mathbf{a}_3$, restoring the periodic boundary condition and yielding $\mathbf{b} \cdot \mathbf{M}_\nu = \pm\pi$.

Fig. 2(a) shows the geometric structure of the considered SDs, whose core extends along the $z$ direction. The calculated band structures in Fig. 2(b) demonstrate the emergence of topological gapless states within the bulk band gap, consistent with the criterion of Eq. (2) [38]. Like the helical topological edge states of QSHIs, the topological SD states contain four (two pairs) 1D modes. These four states are highly localized at the dislocation cores, with each dislocation carrying a pair of oppositely propagating gapless states. In Fig. 2(c), we present the spin textures of one pair of helical modes localized at one SD, showing that the spin polarization whirls along the $k$ path. Consequently, one can effectively manipulate the spin polarization direction of the topological SD states by tuning the Fermi energy, which provides an additional degree of freedom for spintronics applications [42-44]. We also found that the distortion strength δ$t$ is the key factor to change the period of spin polarization evolution (see Supplemental Material [45]), implying that the detailed spin texture of topological SD states is dependent of the structural symmetry, which affords the possibility of delicate manipulation via strain engineering [50].

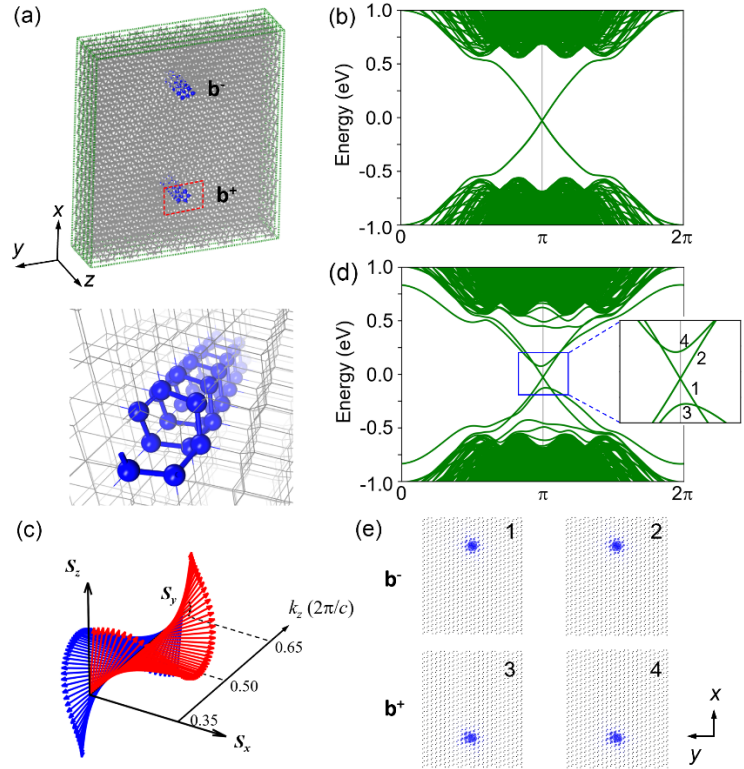

FIG. 2. (a) Top panel: a pair of separated SDs in diamond lattice. The green dashed lines indicate the supercell containing 1248 single-orbital sites, within which a pair of SDs having opposite Burgers vectors (**b**$^+$/**b**$^-$) are created, and the dislocation cores are highlighted by blue. The red rectangle marks the region where the local field is applied. Bottom panel: perspective view of one SD. (b) Energy bands of the dislocated supercell, showing SD-induced topological gapless states. (c) Spin textures of a pair of topological SD states. (d) Energy bands in the presence of local field at one SD. The inset shows the zoomed-in bands near the band crossing point. (e) Charge density distributions of the four bands indicated in (d), localized around **b**$^+$/**b**$^-$ SD core.

We then explore the possibility of tuning the topological SD states by a local Zeeman field, which can be introduced by applying a magnetic field or proximity to a magnetic material, and has been demonstrated efficient in tailoring the topological states [51-53]. Its effect can be included in the TB model by adding the following term, $H_B = \sum_{i \in Z_B} c_i^\dagger \bm{B} \cdot \bm{s} c_i$. Here $Z_B$ indicates a local region where the Zeeman field is applied, as illustrated by the red rectangle in Fig. 2(a). Fig. 2(d) shows the band structures for $B_z = 1$. After turning on a finite $z$-direction Zeeman field, the Kramers degeneracy of one pair of SD states are broken: opening a gap between two topological SD states, while keeping the other two gapless. By analyzing the charge density distributions of the four states as displayed in Fig. 2(e), we identify that the gapped bands are states located around the SD that is covered by the local Zeeman field, as expected, whereas the gapless states are localized around the other SD away from the Zeeman field. The spin textures of the preserved gapless states are also checked to remain the same as before applying the local field. These results indicate that the local Zeeman field can selectively remove one pair of the topological dislocation states, meanwhile leaving the other pair intact. In other words, the application of a local Zeeman field breaks the time-reversal symmetry and destroys the topological helical states, driving the system into another topological nontrivial phase displaying topological chiral states. This essentially realizes a field-controlled switching between topological helical and chiral states. The mechanism of gap opening by a Zeeman field can be found in Supplemental Material [45]. It is worth noting that similar to the proposed selectively gapping of topological SD states, experiments have confirmed the feasibility of gapping out only one surface state of 3D TIs by magnetic doping while keeping the other surface state on the opposite side intact [54,55].

*Robust quantized transport of topological SD states.*–Based on the above insights from TB modeling, we next study the quantum transport property of the topological SD states and the tunability by the local Zeeman field using a two-terminal mesoscopic setup as shown in Fig. 3(a). The scattering region is a finite system built by the supercell of Fig. 2(a) containing a pair of SDs, while two semi-infinite metallic leads are attached along one SD core, the channel, to provide transport carriers as shown in Fig. 1. The Kwant software package [56] is employed to build the device and numerically solve the transport problem within the Landauer-Buttiker formalism [57]. Fig. 3(b) displays the calculated conductance as a function of energy in the absence and presence of the local Zeeman field at the other SD, the "gate". The quantized plateaus are observed for both cases, but the conductance is quantized to $2e^2/h$ and $e^2/h$, respectively, consistent with the number of quantized transmission channels for the 1D topological helical and chiral states. Here the conductance plateau only exists in an energy range of about 40 meV, much smaller when compared with the large nontrivial band gap as shown in Fig. 2(b). This is caused by the finite size of the scattering region in the transverse directions, which gives rise to surface states that drastically reduce the band gap [45].

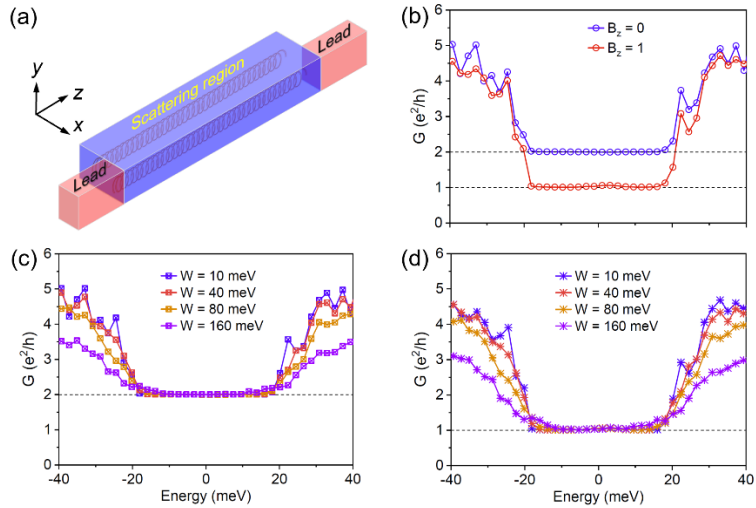

FIG. 3. (a) A schematic view of the device setup to conduct transport simulations. The size of the scattering region is 1×1×80 times of the periodic supercell. The leads cover half of the scattering region and one SD, while the local field is applied on the other SD. (b) Conductance as a function of energy in the absence and presence ($B_z = 1$) of the local Zeeman field. (c) and (d) are conductance results against disorders for $B_z = 0$ and 1 respectively. W indicates the strength of disorder. For a specific disorder strength, the conductance is averaged over 30 different disorder configurations.

We further check the robustness of the topological SD states against disorder. The Anderson-type onsite disorder is considered by including an additional term $W\sum_i \xi_i c_i^\dagger c_i$ to the TB model of the scattering region. Here $\xi_i$ is a random variable uniformly distributed in the interval of [-1, 1], and W represents the strength of disorder. Fig. 3(c) shows the calculated conductance results of the pristine topological SD states under different disorder strength. One can see that, although the conductance drops with increasing disorder strength for energies outside the nontrivial band gap, the conductance plateau of $2e^2/h$ preserves even for a disorder strength up to W = 160 meV, demonstrating the topological protection and robustness of the SD-induced gapless states. Fig. 3(d) shows the conductance results when the local Zeeman field is turned on. The conductance plateau quantized at $e^2/h$ persists for all considered disorder strength, confirming that the gapless states are topologically protected even in the presence of a local Zeeman field. Our transport calculations in clean and disordered regimes verify the topologically protected conductance of the topological SD states, which can be switched from one quantized value ($2e^2/h$) to another ($e^2/h$) using a local Zeeman field as a knob. This finding is very inspiring for disclosing a new recipe to realize two topologically protected conductance signals in a single material, which renders our proposed TFETs with unprecedented high robustness and precision. The two distinct quantized conductance values effectively lend an infinite ON/OFF ratio by signal processing of (2−1)/(1−1), using an exact reference conductance of integer "1".

*Material candidates*.–Before closing, possible materials hosting the predicted tunable topological SD states are explored. Considering the diamond lattice we used for model study, bulk bismuth holding a similar crystal structure naturally comes to our mind [58]. However, the pristine bulk bismuth is a higher-order TI with topological indices (0;000) [59], which cannot give rise to topological dislocation states according to Eq. (2). Even though it can be tuned into a strong TI phase by strain, the bulk bands will overwhelm the topological SD states due to its non-global narrow band gap [60]. To obtain clean topological SD states separated from bulk states, we switch attention to a large energy gap oxide TI, $BaBiO_3$, with cubic perovskite structure (Fig. 4(a)) [39]. $BaBiO_3$ has nonzero topological indices (1;111) and a large nontrivial band gap lies in its conduction bands. Since the topologically relevant band inversion only involves Bi-6$s$ and Bi-6$p$ orbitals [39], one can adopt a simplified model with the basis of Bi-($s$, $p_x$, $p_y$, $p_z$) to capture the essential low-energy physics around its nontrivial band gap (see Supplemental Material for details [45]). Fig. 4(b) and 4(c) plot the corresponding band structures, showing consistence between TB modeling and density-functional-theory calculations. Accordingly, the $BaBiO_3$ structure is reduced to a Bi sublattice as shown in Fig. 4(a). By introducing a pair of SDs to a supercell of the representative Bi sublattice, whose Burgers vectors are **b** = ±**a**$_3$ satisfying Eq. (2) [45], we obtain the band structure in Fig. 4(d). It shows the emergence of topological SD states inside the bulk band gap. We also checked the band structure of the same supercell without SDs, to ensure it does not show gapless states [45].

Next, we investigate the field-tunability of these gapless states by turning on a local Zeeman field $B_z$ covering one of the two SDs. Fig. 4(e) shows that a pair of SD bands open a gap of ~ 20 meV while the other pair remain gapless for $B_z$ = 0.1 eV. Fig. 4(f) displays the corresponding charge density distributions, further confirming that the local field can selectively remove one pair of topological gapless states. Also, the spin textures of the preserved gapless states are shown in Fig. 4(g), which remain intact after applying the local field [45]. Here the spin textures have small y-components, perpendicular to the Zeeman field ($B_z$), resulting in the band gap opening. All these results agree with our Fu-Kane-Mele TB model studies, suggesting BaBiO$_3$ as a potential material candidate to achieve the proposed topological SD states. Considering that topological materials with strong SOC can have very large effective g factors [61], the proposed Zeeman field might be achievable through a feasible magnetic field [45].

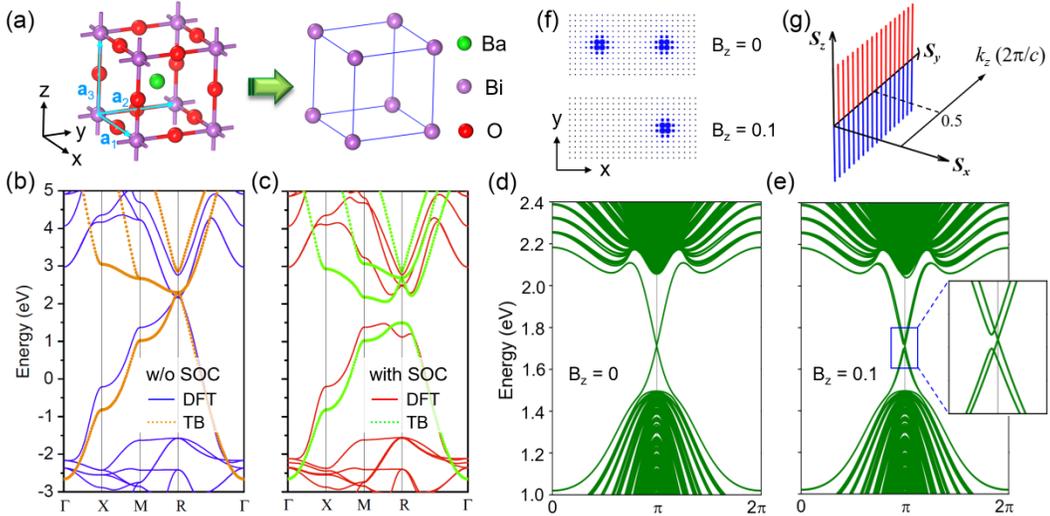

FIG. 4. (a) Atomic structure of cubic BaBiO$_3$ and its Bi sublattice dominating the topological electronic states. (b) and (c) are energy bands of BaBiO$_3$ without and with SOC. (d) and (e) are energy bands of the dislocated supercell in the absence and presence of the local field. (f) Charge

density distributions of the topological gapless states. Here only a portion of the structure around the SDs is shown. (g) Spin textures of the preserved gapless states for $B_z = 0.1$ eV.

*Conclusions.*–The concept of high-fidelity TFETs operating with both ON and OFF states of quantized conductance as we demonstrated is general. It significantly broadens the functionalities of TIs for quantum device applications. The physical principle we disclose for implementing topological dislocation states in FETs may also be applied to edge states of 2D TIs or hinge states of 3D high-order TIs with a proper choice of field direction relative to spin textures; and for device construction, it may be easier to experimentally implement an inner channel of dislocation than an outer channel of edge/hinge on the boundary. Moreover, multiple channels/gates may be fabricated in one material platform by generating dislocation arrays, which could be harder if using edges or hinges. To generalize the proposed mechanism of "fully-topological" ON/OFF signals, we suggest that the Chern-number-tunable quantum anomalous Hall insulators may function as another class of materials to realize the proposed high-fidelity TFET. The Chern number characterizing the dissipationless chiral edge states can be tuned either by varying the composition of the multilayered materials [62-64] or by manipulating the magnetization orientation of specific materials [65], which endows the systems with different quantized conductance. In this context, a feasible "gate" to effectively switch the system between different Chern numbers can be further explored. On a last note, in proximity with a superconductor, our proposed Zeeman field tuned topological dislocation states may offer a new platform for realizing Majorana zero mode at one or many dislocation cores to function as quantum bits.

**Acknowledgement**

We thank X.M. Zhang for helpful discussions. We acknowledge the support by U.S. DOE-BES (Grant No. DE-FG02-04ER46148). Computational resources of this work were supported by CHPC of University of Utah and DOE-NERSC.

Supplemental Material for

**Topological field-effect transistor with quantized ON/OFF conductance of helical/chiral dislocation states**

Xiaoyin Li and Feng Liu*

Department of Materials Science and Engineering, University of Utah, Salt Lake City, Utah 84112, USA

**SI. Methods for first-principles calculations**

The first-principles density-functional-theory calculations are performed by Vienna Ab initio Simulation Package package [1]. The Perdew-Burke-Ernzerhof functional within the generalized gradient approximation [2,3] is used to treat the exchange-correlation interactions and the wavefunction energy cutoff is set to 550 eV. The Brillouin zone is sampled by a k-point mesh with a grid density of $2\pi\times0.04$ Å$^{-1}$ following the Monkhorst-Pack scheme [4]. The energy convergence criterion is $10^{-5}$ eV.

**SII. Spin texture of topological screw dislocation (SD) states with varied model parameters**

The adopted diamond-lattice-based tight-binding (TB) model (Eq. (1) of the main text) has three parameters, the NN hopping strength $t$, the [111]-direction distortion strength $\delta t$ and the SOC strength $\lambda_{SO}$; one can fix $t$ and separately tune $\delta t$ and $\lambda_{SO}$ to numerically study their influence on the detailed spin texture. The results are summarized in Fig. S1. One can see that the distortion strength $\delta t$ is the key factor to change the period of spin polarization evolution.

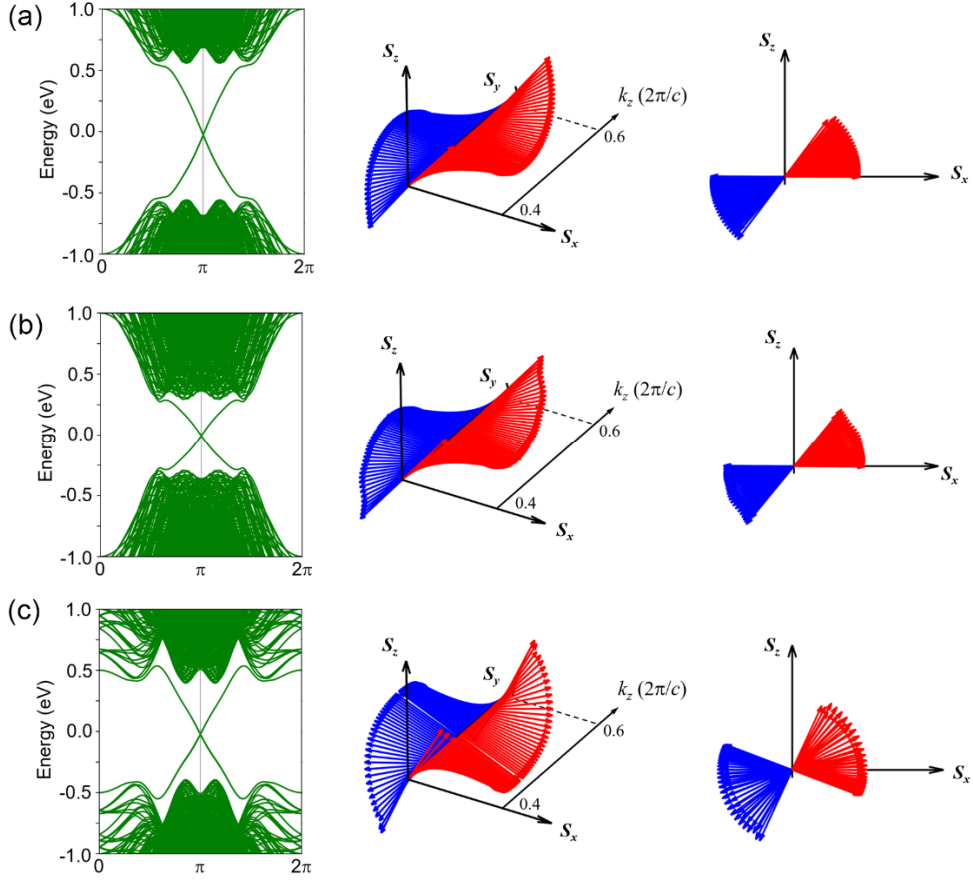

FIG. S1. Band structures (left panels) and spin textures (middle and right panels) of one pair of gapless states for different Fu-Kane-Mele TB model parameters. (a) $t = -1$, $\delta t = -1$, $\lambda_{\text{SO}} = -0.3$. (b) $t = -1$, $\delta t = -1$, $\lambda_{\text{SO}} = -0.15$. (c) $t = -1$, $\delta t = -0.5$, $\lambda_{\text{SO}} = -0.3$. All parameters are in unit of eV. Red and blue arrows in the middle and right panels indicate spin-up and -down directions, respectively.

**SIII. Zeeman-field tuned band gap in 1D dislocation line states**

The mechanism of Zeeman-field-opened band gap in the 1D topological dislocation line states can be elucidated by a low-energy effective Hamiltonian expanded at the high-symmetry band-crossing $k$ point, $H_{\text{eff}} = v_F k \boldsymbol{\sigma}$, with $v_F$ being the Fermi velocity and $\boldsymbol{\sigma}$ being the Pauli matrices describing the spin textures of the two states. For simplicity, we set $v_F = 1$. The band structure of this effective Hamiltonian is shown in Fig. S2(a), reproducing the low-energy physics of one pair of the 1D topological SD states near the band-crossing point. When a local Zeeman field is applied to this pair of states, the effective Hamiltonian is modified to $H'_{\text{eff}} = v_F k \boldsymbol{\sigma} + \mathbf{B} \cdot \boldsymbol{\sigma}$, with the second term representing the Zeeman field and **B** illustrating its direction and strength. Here we fix the strength of the Zeeman field to B = 1 for the convenience of exploring its direction influence on band structures. Fig. S2(b) and (c) show the band structure results under the Zeeman field parallel and perpendicular to the spin textures respectively. One can see that, a Zeeman field **B**//$\sigma$ will not open a gap but only leads to an energy shift. On the contrary, the gapless bands open a gap in the presence of a perpendicular Zeeman field **B**$\perp$$\sigma$ (Fig. S2(c)). Fig. S2(d) displays the band structure evolution when the Zeeman field direction is continuously tuned from parallel (0°) to perpendicular (90°) to spin textures, from which one can see that, for a Zeeman field **B** with arbitrary direction, its $B_\perp$ and $B_{//}$ components collectively modify the bands. The former leads to band gap opening while the latter results in an energy shift.

We also checked the influence of Zeeman field direction on the topological SD states of diamond-lattice-based TB model adopted in the main text, which shows very good agreement with the effective Hamiltonian results. In Fig. S3(a) we plot the details of spin texture for topological SD states near the band-crossing point, which has dominant $x$-component and minor $y$- and $z$-component. Consequently, a $x$-direction ($B_x$) Zeeman field leads to an expected energy shift of the topological SD bands with a small band gap opening (Fig. S3(b)). On the other hand, the $y/z$-direction Zeeman field opens a large band gap and with only a slight energy shift, as displayed in Fig. S3(c)/(d).

We note that the small region of the local field as illustrated by the red rectangle in Fig. 2(a) of the main text is restricted by the size of the supercell that hosts the dislocations. For larger sample where the two dislocations are far apart from each other, one can apply the local field in a larger region as long as it leaves the other pair of topological dislocation states intact, as demonstrated by our calculations as shown in Fig. S4.

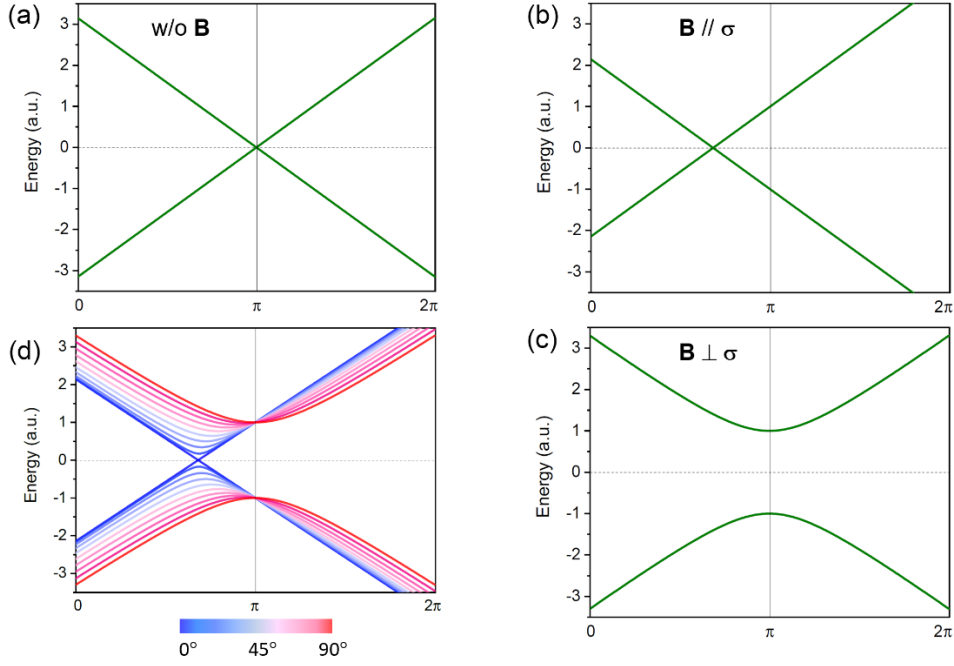

FIG. S2. The Zeeman-field tuned band structures for the low-energy effective Hamiltonian. (a) Band structure without applying Zeeman field. (b) and (c) Band structure in the presence of the Zeeman field, which is parallel and perpendicular to the spin textures respectively. (d) Band structure evolutions with the Zeeman field tuned from parallel (0°) to perpendicular (90°) direction.

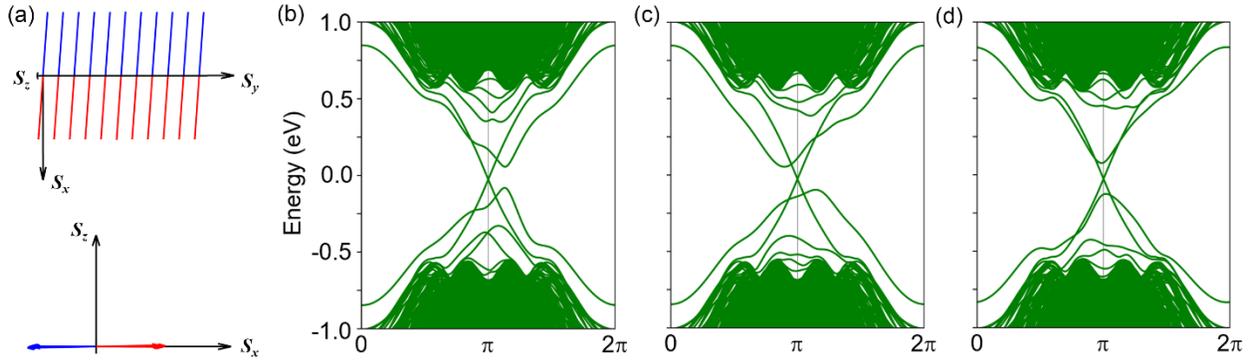

FIG. S3. The local Zeeman-field tuned band structures of dislocated diamond-lattice supercell as in Fig. 2 of the main text. (a) The *x-y* and *x-z* plane views of the spin textures of topological screw dislocation states near the band-crossing point. (b)-(d) Band structure in the presence of the local Zeeman field, with **B** along the *x* [$\mathbf{B} = (B_x, 0, 0)$], *y* [$\mathbf{B} = (0, B_y, 0)$] and *z* [$\mathbf{B} = (0, 0, B_z)$] direction respectively. The strength of the Zeeman field is 1 eV as in the main text.

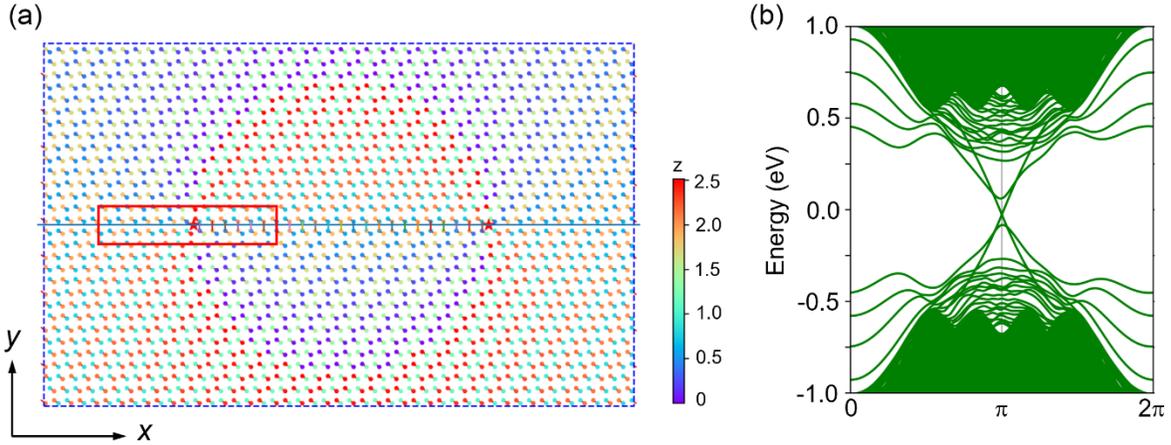

FIG. S4. The local Zeeman-field tuned band structures for a large supercell containing 2760 single-orbital sites. (a) Top view of the supercell structure with a pair of screw dislocations. The stars and the red rectangle represent the dislocation cores and the region where the local field is applied, respectively. The color bar indicates the $z$-direction coordinates in unit of Å. (b) Energy bands of the dislocated supercell in the presence of the local field $B_z = 1$ eV.

## SIV. Finite-size induced trivial surface states

The 2D surfaces of 3D topological insulators (TIs) have gapless band structures. However, when the TI acts as the 1D channel for quantum transport simulations, some surfaces are cut into finite size and the gapless states from these surfaces will interact and open a gap at the hinge where different surfaces meet. The resultant band gap is much smaller than the bulk band gap, leading to the reduced band gap size of the 1D sample (as shown in Fig. S5) compared with the periodic case (Fig. 2(b) and (c) of the main text).

Nevertheless, if these finite-size induced trivial surface states are suppressed from participating in transport, one can observe a conductance plateau associated with the topological SD states in an energy window comparable with the nontrivial band gap of Fig. 2(b). Practically, this can be achieved by introducing strong disorders to the surface region responsible for these undesired states as demonstrated in Fig. S6. Additionally, one can use local leads only covering the central region where SDs reside to get rid of the trivial surface states since these states are distributed at the surfaces and have a finite penetration length toward the bulk.

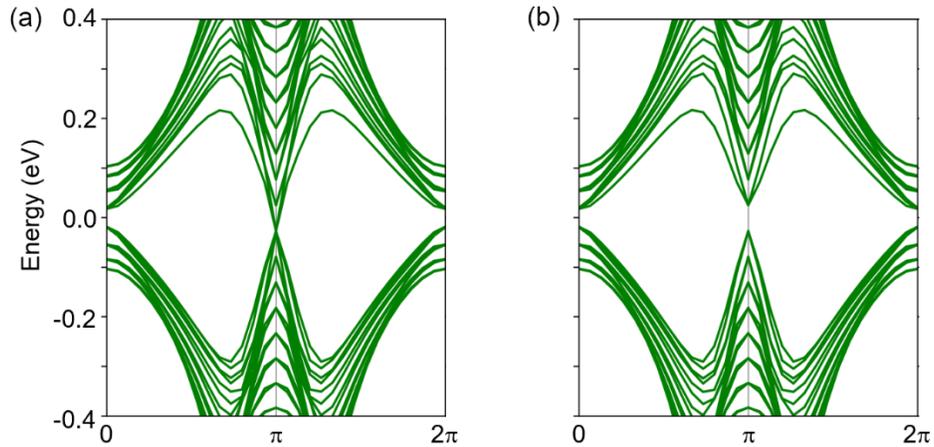

FIG. S5. Band structures of the supercell when open boundary condition is adopted along the $x$ and $y$ directions. The supercell structure is the same as that in Fig. 2(a) of the main text, which has 1248 orbitals per supercell. (a) Band structures when the supercell contains a pair of screw dislocations. (b) Band structures for the same supercell without screw dislocations. One can see that the nontrivial band gap is significantly reduced by the finite-size induced states.

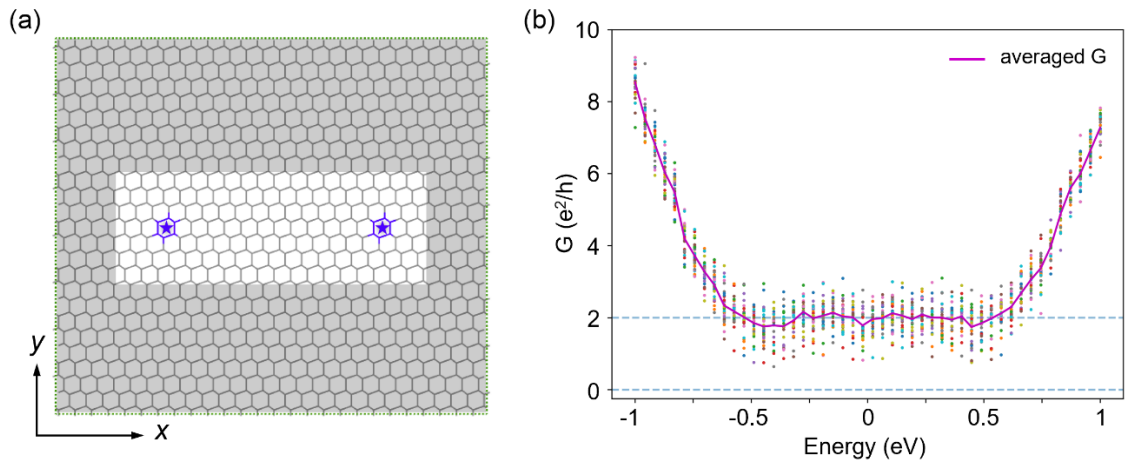

FIG. S6. Conductance of the topological screw dislocation states when the finite-size induced states are suppressed by strong disorder. (a) Cross-section view of the scattering region for the two-terminal device. The gray-shaded region indicates the region where Anderson-type disorder is introduced. (b) The calculated conductance in the presence of strong disorder. The dots represent conductance results for different configurations of disorder and the solid line is the averaged

conductance over 30 configurations. The disorder effect is included by adding an additional term $W\sum_i \xi_i c_i^\dagger c_i$ to the TB model of Eq. (1) in the main text. Here $\xi_i$ is a random variable uniformly distributed in an interval of [-1, 1], and W represents strength of the disorder. The adopted strong disorder strength is set at W = 20 eV, which is sufficient to completely localize the finite-size induced states. Here the conductance plateau exists in a large energy range in agreement with the nontrivial band gap in Fig. 2(b) of the main text.

## SV. Tight-binding model for the real material candidate BaBiO$_3$

The TB model for BaBiO$_3$ with four orbitals ($s$, $p_x$, $p_y$, $p_z$) per site reads

$$H = \sum_{i\alpha} \varepsilon_\alpha c_{i\alpha}^\dagger c_{i\alpha} + \sum_{\langle i\alpha, j\beta \rangle} t_{i\alpha,j\beta} c_{i\alpha}^\dagger c_{j\beta} + i\lambda \sum_i \left( c_{ip_y}^\dagger \sigma_z c_{ip_x} - c_{ip_x}^\dagger \sigma_z c_{ip_y} \right). \quad (S1)$$

Here $i$, $j$ and $\alpha$, $\beta$ are site and orbital indices respectively. $\varepsilon_\alpha$ in the first term is the onsite energy of orbital $\alpha$. The second term contains the NN hoppings which are parameterized following the Slater-Koster scheme. The third term represents the onsite SOC of atomic $p$ orbitals with the strength of $\lambda$. By fitting the model TB band structures to density-functional-theory bands as shown in Fig. 4(b) and 4(c) of the main text, the parameters (in unit of eV) are extracted to be $\varepsilon_s = 0.10$, $\varepsilon_p = 4.36$, $t_{ss\sigma} = -0.460$, $t_{sp\sigma} = 0.675$, $t_{pp\sigma} = 0.844$, $t_{pp\pi} = 0.094$, and $\lambda = 0.4$. To further validate this model, the (001) surface states are calculated as shown in Fig. S7, which reproduce well the previous results.

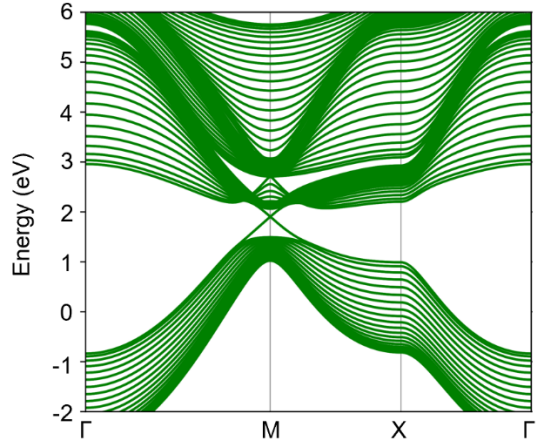

FIG. S7. (001) surface states of the constructed Slater-Koster tight-binding model for BaBiO$_3$, reproducing the result of previous study with the Dirac point located at the M point. The slab thickness used is 20 layers.

SVI. Supplemental figures for the real material candidate BaBiO$_3$

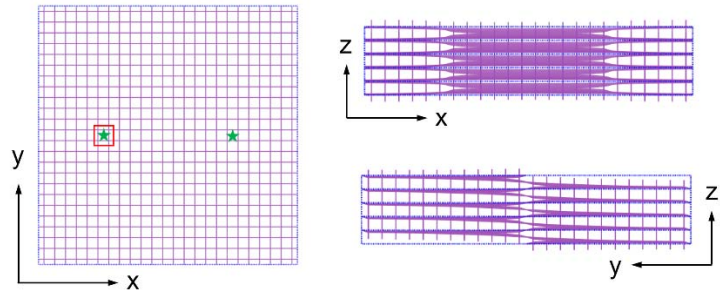

FIG. S8. Top and side views of dislocations in the simplified cubic lattice for BaBiO$_3$, where green stars and red rectangle indicate the dislocation cores and local Zeeman field region respectively. The supercell contains 576 sites and 2304 orbitals.

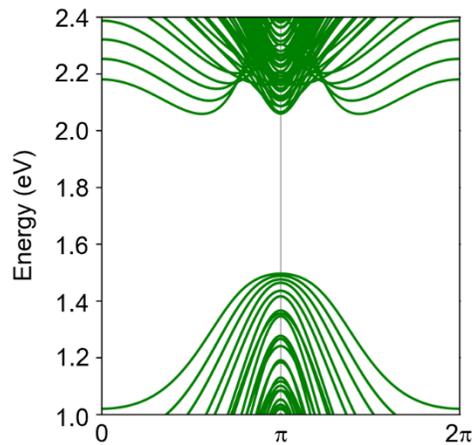

FIG. S9. Band structures of the supercell as shown in Fig. S8 but without screw dislocations.

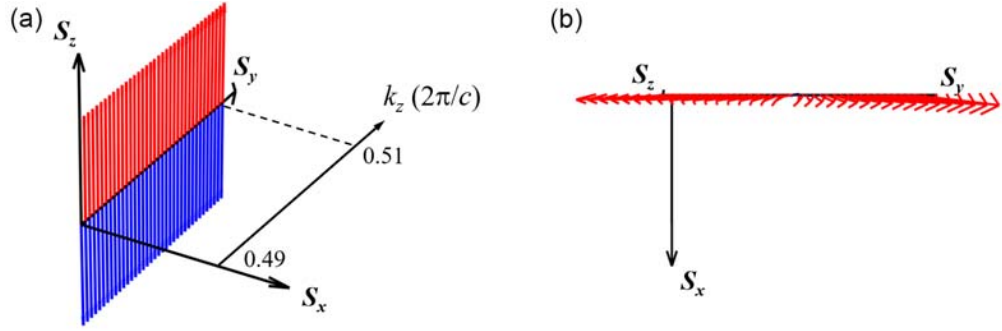

FIG. S10. Spin textures of one pair of gapless states for the bands in Fig. 4(d) of the main text. (a) A perspective view of the spin textures. (b) The *x-y* plane view of the spin textures, showing dominant *z*- and minor *y*-components. The former is parallel to the *z*-direction local Zeeman field and contributes an obvious energy shift, while the latter is perpendicular to the local Zeeman field and opens a band gap; collectively, they lead to the band structure as shown in Fig. 4(e) of the main text.

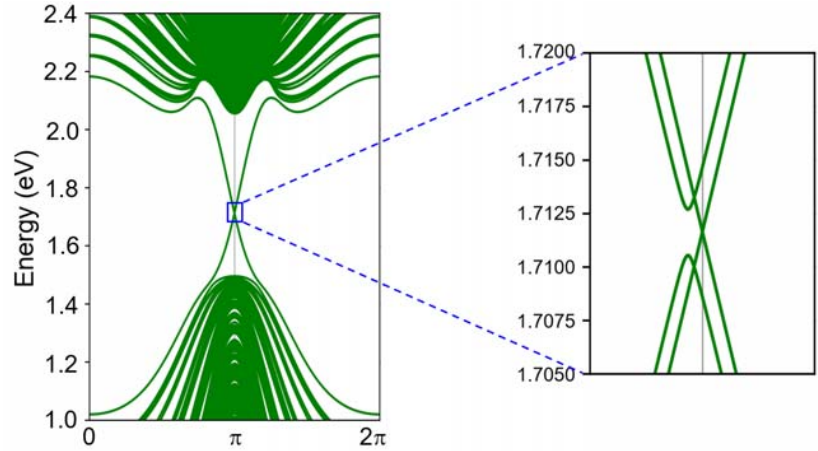

FIG. S11. Band structures of the dislocated supercell as shown in Fig. S8 under a smaller local Zeeman field $B_z = 0.01$ eV, which opens a band gap of ~ 2.5 meV for one pair of topological dislocation states. The corresponding magnetic field strength should be $B_z/g\mu_B$, with $g$ and $\mu_B$ being the effective $g$ factor and Bohr magneton respectively. The required magnetic field is estimated 8.6 T if one assumes an effective $g$ factor of 20.